\documentclass{article}
\usepackage{fleqn}
\usepackage{amsmath,amssymb}

\newcommand{\const}{{\mbox Const}}
\begin{document}

\begin{center}
{\bf\large The complete relativistic kinetic model of violation of
symmetry in isotopic expanding plasma and production of baryons in
hot Universe. I. Exact model.}\\[12pt]
Yu.G.Ignat'ev, K.Alsmadi\\
Kazan State Pedagogical University,\\ Mezhlauk str., 1, Kazan
420021, Russia
\end{center}

\begin{abstract}
The complete model of production of baryons in expanding
primordial symmetrical hot Universe is constructed in terms of
general relativistic kinetic theory.\end{abstract}

\section{Introduction}
An important example of local thermodynamical equ\-i\-li\-bri\-um
(LTE) breaking by massive particles during the cosmological
expanding is process of generation of baryons (baryogenesis) in
primordial baryon - sym\-met\-ri\-cal Universe. On \cite{Sahar},
\cite{Kuzmin} there was promoted a hypothesis that observing
baryon skewness of Universe
\begin{equation}\label{0}
\delta_b=\frac{n_B}{n_\gamma}\sim 10^{-9}
\end{equation}
is caused by $CP$- non - invariant processes, which break the
conservation of baryon's charge. An example of such process is
decay of superheavy $X$- bosons, to which Grand Unified Theories
(GUT's) adduce:\footnote{Here and further $\bar{a}$ means
antiparticle $a$. }
\begin{equation}\label{1} X \rightleftarrows
\bar{q}+\bar{q};\:\: X\leftrightarrows q+l,
\end{equation}
where $q$ - quark's symbol, $l$ - lepton's symbol. In \cite{Okun},
however is shown that if given processes are run in LTE's
conditions then even the existence of ба - non-invariance and the
breaking of baryon charge's conservation is insufficiently for
appearance of baryon's excess over anti-baryons. For production of
the baryon's charge in primordial baryon-symmetrical Universe
be\-si\-des stated facts require the breaking of LTE in reactions
of such type (\ref{1}) and withdrawal of  baryons from statical
equilibrium's conditions. Such possibility: %
\begin{equation}\label{2_0}\tau_x \gtrsim
t\end{equation} %
($\tau_x$ - time of half-decay of $X$ - boson, $t$ - cosmological
time ) is realized under condition \cite{OtheIgnat}:
\begin{equation}\label{2}
m_X>\alpha_X m_{pl}\sqrt{N}, \end{equation}
($m_X$ - mass of
$X$-boson, $m_{pl}=\sqrt{\frac{\hslash c}{G}}$ - Planck's mass,
$N$ - number of degrees of freedom). Condition (\ref{2})
stringently limits from below allowable values of  $X$- bosons
masses.

In Ref. \cite{OtheIgnat}, \cite{Weinberg1} (see too
\cite{Weinberg2}, \cite{Dolgov} and etc.) there were made
estimates of baryon skewness of Universe with the assumption of
LTE's breaking. For attitude of baryons number's density to
density of total entropy, $\mathcal{S}$, there was obtained:
\begin{equation}\label{3}
\delta_\mathcal{S}=\frac{n_b}{S}\simeq \frac{45
\zeta(3)}{4\pi^4}\frac{N_X}{N} \Delta r, \end{equation} where
$\zeta(x)$ - $\zeta$ - Rieman's function, $N_X$ - the number of
types of superheavy  $X$- bosons, $\Delta r$ - difference of
comparative probabilities of decays in channels
$$X\rightarrow q+l\: \mbox{ш}\: \bar{X}\rightarrow
\bar{q}+\bar{l},$$ which appears in consequence of breaking of
$CP$ - invariance. Further in papers \cite {Fry1}-\cite{Fry3} on
basis of relativistic kinetic theory there were made numerical
accounts of attitude $n_B/S$, which basically confirmed foregoing
estimates. In cited earlier papers on the basis of made
calculations there was found the inferior limit of $X$ - boson's
mass:

\begin{equation}\label{3.1}
m_X\geq 10^{16}\mbox{Gev},\end{equation} which created
considerable difficulties for standard $SU(5)$ theory.

Previously mentioned papers however have con\-si\-de\-rab\-le
deficiency - instead of direct solution of kinetic equations for
$X$ - bosons under conditions of {\it considerable} LTE's breaking
boson functions of distribution are described by quasi-equilibrium
(Bol\-zman's) di\-s\-tri\-bu\-ti\-ons, which parameters are
defined from hydrodynamical equations for moments of
di\-s\-tri\-bu\-ti\-on's function, i.e., in this papers actually
is used hydrodynamical method of Grad. According to results of
relativistic kinetic theory (see for example,
\cite{Ignat1}-\cite{Ignat3}), global thermodynamical equilibrium
in homogeneous isotropic expanding plasma is reached only in
extreme non-relativistic limit, or in extreme ultrarelativistic
limit. In range of intermediate energies of particles and on
conditions that LTE is breaking, the distribution of particles is
not approximated by equilibrium distribution. In Ref.
\cite{Ignat4_0}-\cite{Ignat4} of one of authors in the framework
of kinetic theory there was found non-equilibrium function of
distribution of $X$ - bosons and shown that this function may
essentially differs from equilibrium function. Therefore results
obtained in \cite{Fry1}-\cite{Fry3}are correct if  {\it strong}
inequality (\ref{2}) is applied, but needed a correction in other
areas. The estimate (\ref{3}) although seems to appear
sufficiently accurate, but doesn't describe such situations, when
condition (\ref{2}) is not realized while width of covering of
experimental and theoretical im\-p\-li\-ca\-ti\-ons $\delta_S$
permits possibility when baryon charge is produce in conditions,
that are less favorable than (\ref{2}).

The answer to the question, what kind will be magnitude $\delta_S$
at $m_X\leq\alpha_X m_{pl}\sqrt{N}$, could be given only by
detailed kinetic analysis. On the assumption of confidence to the
hypothesis of primordial baryon- and charge- symmetrical Universe,
such analysis from the other side allows more definitely indicate
range of possible implications of fundamental constants of Grand
Unified Theories. In the middle of 80-s - beginning of 90-s by one
of authors was formulated the kinetic model of describing of
symmetry's breaking processes and there were obtained some
estimations, following from this model (Ref.
\cite{Ignat4_0}-\cite{Ignat5}). Specifically on basis of obtained
estimations there was formulated a hypothesis that accounting of
kinetic of baryogenesis process will permit to reduce the low
estimation of masses of superheavy bosons to quantity of order
$5\cdot 10^{14}$Gev. However there wasn't executed detailed
analysis of this model in this papers and on account of external,
researches of this problem weren't complete. The pur\-po\-se of
this paper is exactly carrying out of more detailed analysis of
the kinetic model of baryogenesis and co\-n\-struc\-ti\-on of the
numerous model of considered events. Let us notice that though
this paper contains results for concrete model of interactions
based on the minimal $SU(5)$ symmetry, the generalization of these
results for other analogical field theories isn't hard - it is
reduce to arithmetical recalculation of corresponding
coefficients.

\section{The algebra of interactions}
Let us consider for example standard $SU(5)$ model of interactions
(see for example \cite{Dolgov}). There are involve 12 vectorial
calibrating bosons which represent themselves 2 charged color
triplets in this model:
$$\{X^i_{v,-4/3},\: \bar{X}^i_{v,4/3};\: X^i_{v,-1/3},
\bar{X}^i_{v,1/3}\}$$ - $i$ - color index (red, green, blue), the
subscript - electrical charge - and 12 scalar Higgs's bosons:
$$\{X^i_{s,-1/3},\: \bar{X}^i_{s,1/3};\: X^i_{s,-4/3},
\bar{X}^i_{s,4/3}\}.$$

In more common models the number of vectorial bosons reduces to
24, - and 2 more triplets add at the same time
$$\{X^i_{v',-1/3},\:\bar{X}^i_{v',1/3};\:X^i_{v',2/3},\bar{X}^i_{v',-2/3}\},$$
and the number of Higgs's bosons reduces to 30, and at that add 3
$SU(3)$ - triplets, which are included in 3 $SU(2)$ triplets:
$$\{X^i_{s',-1/3},\:
\bar{X}^i_{s',1/3};\:X^i_{s',2/3},\:\bar{X}^i_{s'-2/3}; \:
X^i_{s',-4/3},\: \bar{X}^i_{s',4/3}\}.$$ Superheavy $X$-bosons are
often calling $X,Y,Z$-bosons according to their charges:
$-4/3,-1/3,2/3$. The electric charge's conservation laws allow to
run only these reactions of decay/creation of these bosons:
\begin{equation}\label{4}
\begin{array}{l}
X_{-4/3}\rightleftarrows \bar{q}_\alpha + \bar{q}_\alpha;\:X_{-4/3}\rightleftarrows q_\kappa+l_e;\\
\\
X_{-1/3}\rightleftarrows \bar{q}_\alpha+\bar{q}_\kappa;\:
X_{-1/3}\rightleftarrows q_\kappa+l_\nu;\: X_{-1/3}\rightleftarrows q_\alpha+l_e;\\
\\
X_{2/3} \rightleftarrows \bar{q}_\kappa +
\bar{q}_\kappa;\:X_{2/3}\rightleftarrows q_\alpha + l_\nu\\
\end{array}
\end{equation}
where $\alpha$ - index of apoquark ($\alpha=u,c,t$ - quarks with
charge $2/3$), $\kappa$ - katoquark ($\kappa=d,s,b$ quarks with
charge $-1/3$), $e$ - symbol of charged lepton ($e=e,\mu,\tau_-$,
charge=-1), $\nu$ - symbol of neutral lepton
($\nu=\nu_e,\nu_\mu,\nu_\tau$). In reactions (\ref{4}) color
charge is also conserving so therefore in reactions of decay of
$X$-boson and in reaction of quark annihilation antiquarks of
different colors, additional to $X$ -boson's color, take part in.
For example:
$$\begin{array}{l}
\bar{X}^{R}_{-4/3}=d^{R}_{r}+e_-=\bar{u}^{\bar{G}}_{l}+\bar{u}^{\bar{B}}_{l};\\
\\
\bar{X}^{G}_{-4/3}=d^{G}_{l}+e_-=\bar{u}^{\bar{R}}_{l}+\bar{u}^{\bar{B}}_{l},\\
\end{array}
$$
where $R,G,B$ - indications of red, green, blue colors;
corresponding overlined indexes $\bar{R},\bar{G},\bar{B}$ -
anticolors, which are equal to sums of corresponding additionals
colors, $l,r$ - indications of left and right particles. We may
suppose, as it often doing that in right parts of reactions
(\ref{4}) particles only with the same charm take part in. In that
case in standard variant $SU(5)$ there will be 90 couples of
reactions of type (\ref{4}), and in expended variant of theory -
198 couples\footnote{With account of decay and creating of
superheavy antibosons.}. Then, the number of degrees of freedom in
standard variant of theory is equal:
$$N=\sum\limits_B (2s+1) +\frac{7}{8}\sum\limits_F (2s+1)=185$$
- 24 vectorial bosons ($s=1$), - 29 scalar bosons ($s=0$), 36
quarks ($s=1/2$), 12 leptons ($s=1/2$). In expanded variants
$SU(5)$ this number is even greater.

\section{Kinetic equations}
Let us for convenience express averaged on spin states invariant
elements of transition matrix of decays of superheavy bosons
with the help of nondimensional numbers $(r_i,\bar{r}_i)$:%
\begin{equation}\label{5}
\begin{array}{ll}
\left|M_{X\rightarrow ql_e}\right|^2= \frac{1}{3} s^2r_e; &
\left|M_{X\rightarrow ql_\nu}\right|^2= \frac{1}{3} s^2r_\nu; \\
&\\
\left|M_{\bar{X}\rightarrow \bar{q}\bar{l}_e}\right|^2=
\frac{1}{3}s^2\bar{r}_e; & \left|M_{\bar{X}\rightarrow
\bar{q}\bar{l}_\nu}\right|^2= \frac{1}{3}s^2\bar{r}_\nu;\\
&\\
 \left|M_{X\rightarrow \bar{q}\bar{q}}\right|^2=
\frac{1}{6}s^2(1-r); & \left|M_{\bar{X}\rightarrow qq}\right|^2=
\frac{1}{6}s^2(1-\bar{r}),\\
\end{array}
\end{equation}
($r=r_e+r_\nu, \bar{r}=\bar{r}_e+\bar{r}_\nu$), which should be
also provide by indexes of electrical and color charges as well as
by charms of quarks and leptons.

Through $CPT$ - invariance averaged matric elements of reverse
transition (annihilation) are equal:
\begin{equation}\label{6}
\begin{array}{ll}
\left|M_{ql_e\rightarrow X}\right|^2= \frac{1}{3} s^2\bar{r}_e; &
\left|M_{ ql_\nu\rightarrow X}\right|^2= \frac{1}{3} s^2\bar{r}_\nu;\\
&\\
\left|M_{\bar{q}\bar{l}_e\rightarrow \bar{X}}\right|^2=
\frac{1}{3}s^2r_e; & \left|M_{\bar{q}\bar{l}_\nu\rightarrow
\bar{X}}\right|^2= \frac{1}{3}s^2r_\nu;\\
&\\
\left|M_{\bar{q}\bar{q}\rightarrow X}\right|^2=
\frac{1}{6}s^2(1-\bar{r}); & \left|M_{qq\rightarrow
\bar{X}}\right|^2= \frac{1}{6}s^2(1-r).\\
\end{array}
\end{equation}

Invariant matric elements of double-particle decay's probability
are constant. Calculation of total probability of decay in
one-loop approximation in the context of, for example, standard
$SU(5)$ model gives:

$$W_{X\rightarrow
ql}=\frac{\pi\alpha}{108}(2\pi)^4m^2_X\delta^{(4)}(P_F-P_I),$$
where $P$-total 4-momentums of initial and final
co\-n\-di\-ti\-ons $\alpha\approx 1/45$. In that case:
\begin{equation}\label{7}
\sum\limits_{A,C} s^2=\frac{8\pi m^2_X}{27},\end{equation} %
where summation is carries over all charms and colors of particles
participating in decay of given type of $X$ -boson.

In further however we shall not define concretely $s^2$, supposing
only:
\begin{equation}\label{7a}
s^2=\const.\end{equation} Besides first order reactions (\ref{4})
reactions of higher order, running with participation of
superheavy bosons and breaking $CP$-invariance and also
conservation laws of baryon and lepton charges also are possible.
For example
\begin{equation}\label{8} qq'\stackrel{X}{\rightarrow}
\bar{q}l;\:\: q\bar{l}\stackrel{X}{\rightarrow} \bar{q}\bar{q}'.
\end{equation}

Let us consider further metric of homogenous isotropic Universe
\footnote{Here and further we choose system of units
$\hslash=c=G=1$.}:
\begin{equation}\label{8a}
ds^2=dt^2-a^2(t)(dx^2+dy^2+dz^2).\end{equation}

Let further  $p^i$ - 4-momentum of a particle,
cor\-re\-s\-pon\-ding to relation of normalization:
\begin{equation}\label{norm}
(p,p)=(p^4)^2-p^2=m^2 \Rightarrow E=\sqrt{m^2+p^2},
\end{equation}
where  $p^2=a^2(t)((p^1)^2+(p^2)^2+(p^3)^2)$ - square of kinematic
momentum of particle, $E=p^4$ - kinetic energy of particles.

In Ref. \cite{Ignat6} was shown that relativistic kinetic
equations are asymptotically conformally invariant in
ultrarelativistic limit:
\begin{equation}\label{ultralimit}
\frac{<p^2>}{m^2} \rightarrow \infty
\end{equation}
to within $O(m^2/p^2)$. Thus in ul\-tra\-re\-la\-ti\-vis\-tic
limit (\ref{ultralimit}) when energies of all particles
participating in reactions lot more than their rest masses,
kinetic equations in Friedman's metric after corresponding
conformal conversion coincide with kinetic equations in Minkovsky
space, therefore homogenous isotropic expanding of
ul\-tra\-re\-la\-ti\-vis\-tic plasma doesn't lead out from
condition of {\it global} thermodynamical equilibrium. Further in
opposite non-relativistic limit:
\begin{equation}\label{nonrelativistic}
\frac{<p^2>}{m^2} \rightarrow 0,
\end{equation}
when kinetic energies of all particles participating in process
far less than their rest masses, global ther\-mo\-dy\-na\-mi\-cal
equilibrium is again restores in plasma.
\cite{Magdal},\cite{Ignat1}.

Let's introduce according to \cite{Ignat6} conformal mo\-men\-tums
of particles $\mathbb{P,Q}$, which are integrals of motion
relative to metric (\ref{8a}) according to formula:
\begin{equation}\label{8b}
\mathbb{P}=a(t)p\hskip 24pt(=\sqrt{(p_1)^2+(p_2)^2+(p_3)^2}),
\end{equation}
conformal kinetic energy of particles $\mathbb{E}=a(t)E $:
\begin{equation}\label{8c}
\mathbb{E}=\sqrt{a^2(t)m^2+\mathbb{P}^2}
\end{equation}
and corresponding {\it macroscopic conformal parameters of plasma
}, which conserve constant at the ul\-tra\-re\-la\-ti\-vis\-tic
stage of expanding:
\begin{align}\label{conforT}
\mathcal{T}= a(t) T &\:-&  \text{temperature};\\
\mathcal{N}= a^3(t) n &\:-&  \text{number density of particles};\label{conforn}\\
\mathcal{E}=a^4(t) \varepsilon &\:-& \text{energy density}.
\label{conforE}
\end{align}

Kinetic equations for plasma in expanding ho\-mo\-ge\-no\-us
Universe with participation of $X$ -bosons we will write in
symbolic form \cite{Ignat5}:

\begin{eqnarray}\label{9}
\frac{1}{a}\sqrt{a^2m^2_X+\mathbb{P}^2}\frac{\partial
f_X}{\partial t}=\sum
I_{X\rightarrow \bar{q}\bar{q}'} +\sum I_{X\rightarrow ql};\\
\frac{1}{a}\sqrt{a^2m^2_X+\mathbb{Q}^2}\frac{\partial
f_F}{\partial t}= \sum I_{X\rightarrow FF'} +
\sum^{(2)}I_{FF'\rightarrow\ldots}\nonumber\\\label{10} +
(\text{other interactions}),
\end{eqnarray}
where $f_a(P,t)$ - phase-space densities, $F$ - index of fermions,
$\mathbb{P}$, $\mathbb{Q}$ - conformal momentums of $X$ -bosons
and fermions, correspondingly. Expression ''other
in\-te\-ra\-c\-ti\-ons'' in (\ref{10}) means integral of
collisions for all other interactions which don't violate
conservation of baryon and lepton charges: quark-gluon, lepton and
quark decays of $W$- ш $X$-bosons, annihilation etc. All these
reactions run in conditions of LTE and lead to establishment of
thermal equilibrium of quarks and leptons. Integrals of
interactions of particles we will write in relativistic-invariant
form \cite{Ignat5a}:

$$
I_a(x,P_a)=\hskip
5cm$$\begin{equation}\label{17}-\sum\limits_{f,i} \int
\delta^{(4)}(P_f-P_i)
(Z_{fi}W_{if}-Z_{if}W_{if})\underset{f,i}{\Pi'} d\pi,
\end{equation} where summation is carrying out by all initial, $i$, and final,
$f$, conditions of particles, integration ($d\pi'$) is carrying
out by all four-dimensional momentum spaces, except space of
particle of $a$- sort, $W_{fi}$ ш $W_{if}$ - corresponding
invariant scattering matrixes which relate with invariant
amplitudes of scattering $M_{if}$ by formulas \cite{QuantField}:

\begin{equation}\label{17a}
W_{fi}=(2\pi)^4|M_{fi}|^2\cdot 2^{-(N_i +N_f)}-
\end{equation}
- $N_i$ ш $N_f$ - numbers of particles in initial and final
conditions. In formulas (\ref{17}) introduced products of initial
and final conditions:

\begin{eqnarray}\label{18}
Z_{fi}=\prod\limits_f f_f\prod\limits_i (1\pm f_i);\nonumber\\
Z_{if}=\prod\limits_i f_i\prod\limits_f (1\pm f_f),
\end{eqnarray}
where symbol ``$-$'' relates to fermions, ``$+$'' - to bosons.
Invariant elements of momentum space volumes are equal:
$$d\pi_a={ \displaystyle\frac{\rho_a\sqrt{-g} d^4P_a}
{(2\pi \hslash)^3}} \delta((P_a,P_a)- \frac{m^2_ac^2}{2}),$$ where
$\rho_a$ - statistical degeneracy factor.

Further, since reactions (\ref{4}), (\ref{8}) run under very high
temperatures $T\gtrsim 10^{14}$Gev, it is possible with very high
accuracy grade to mean that all fermions are ultrarelativistic,
therefore:
\begin{equation}\label{11}
f_{F_a}=\left[\exp(-\lambda_a +\mathbb{Q}/\mathcal{T}
)+1\right]^{-1},
\end{equation}
where conformal temperature $\mathcal{T}=aT,$ is constant in
ultrarelativistic equilibrium plasma, and {\it adduced} chemical
potentials
$$\lambda_a=\frac{\mu_a}{T}$$ are satisfy to series of conditions
of chemical equilibrium, corresponding to reactions executing in
conditions of LTE. Combining algebraic equations corresponding to
these conditions we may come to such conclusion: chemical
potentials of each type of quark aren't depend on their own color
and charm. In that way it remains only 4 independent chemical
potentials which for sim\-pli\-ci\-ty sake, following
\cite{Dolgov}, we will denote by symbols of particles themselves -
$u,d, e, \nu$:

\begin{equation}\label{12}
u=\lambda_{q_\alpha}; \:\: d=\lambda_{q_\kappa}; \:\:
e=\lambda_{l_e};\:\: \nu=\lambda_{l_\nu},\end{equation} for them
at that as for ultrarelativistic particles in conditions of LTE
applies known condition of an\-ti\-sym\-met\-ry:
\begin{equation}\label{12a}
\bar{\lambda}_a=-\lambda_a. \end{equation} In situation when rest
mass of neutral leptons is equal to zero, their chemical potential
is also equal to zero: $\nu=0$, as chemical potential of massless
particles being in thermodynamical equilibrium \footnote{see for
example, \cite{LandauStat}.}. In that case remain only three
independent variables: $u,d,e$.

\subsection{Kinematic equations for fermions}
At completion of chemical equilibrium conditions the last
component in right part (\ref{10}) becomes a zero. The second
component in right part (\ref{10}) describes all reactions
executing with violence of $CP$-invariance, in which quarks,
leptons and virtual $X$-bosons participate. Since these integrals
completely defined by equilibrium functions of distribution of
quarks, leptons and other particles, then using {\it functional
Bolzman's equations} \cite{Ignat7}

\begin{equation}
Z_{fi}-Z_{if}=0\: \Rightarrow\: \sum\limits_{i}\lambda
-\sum\limits_f \lambda'=0,
\end{equation}
Fermion integrals of interaction can be write down in the form of:
\begin{eqnarray}\label{12b}
\sum I_{FF'\rightarrow\ldots}= -\sum \int d\pi_F d\pi_F
d\pi_{F'}f_F f_{F'}\times\nonumber\\
\times \int \prod d\pi_k (1\pm
f_k)(W_{FF'\rightarrow\ldots}-\bar{W}_{\bar{F}\bar{F}'\rightarrow\ldots}).
\end{eqnarray}

There is no need to concretize probabilities of many-particle
conversions $W$ ш $\bar{W}$ for calculation of these integrals  -
it is enough to use optical theorem which is consequence of
unitarity of $S$-matrix, \cite{Dolgov}, \cite{Pilkun}:

\begin{equation}\label{13}
\sum\limits_k \int \prod d\pi_k (1\pm f_k)(W_{if}-W_{fi})=0,
\end{equation}
where summation is carrying out by all {\it final} conditions of
reactions $FF'\rightarrow\ldots$. So in consequence of optical
theorem we will obtain equality:

\begin{eqnarray}
{\sum\limits_{X,F'}}'(1+f_X)(W_{X\rightarrow
\bar{F}\bar{F}}-W_{\bar{X}\rightarrow FF'})d\pi_k +\nonumber\\
{\sum\limits_{k}}'(1\pm f_k)(W_{FF'\rightarrow\ldots
}-W_{\bar{F}\bar{F}'\rightarrow \ldots})d\pi_k =0.\\
\end{eqnarray}
Using this result in equations (\ref{10}) and integrating them by
momentum space of fermions we will receive equations which in
further will be used for definition of chemical potentials of
fermions:%
\begin{eqnarray}\label{14}
\frac{d \mathcal{N}_F}{d t}=a(t){\sum\limits_{\bar{X},F'}}'\int
W_{X\rightarrow FF'}d\pi_X d\pi_F d\pi_{F'}\times \nonumber\\
\left[f_X (1-f_F)(1-f_{F'})-(1+f_X)f_F f_{F'}\right].
\end{eqnarray}

Last equations can be still simplified by accounting of
correlations which are correct to equilibrium fermion distribution
functions (\ref{11}):

$$
\int d\pi_F d\pi_{F'} \delta^{(4)}(p_X-p_F-p_{F'})f_F f_{F'} $$
\begin{equation}\label{15a}
=\frac{1}{2^3\pi^5}f^0_X(F+F')\beta(F,F');\end{equation}
$$\int d\pi_F d\pi_{F'}\delta^{(4)}(p_X-p_F-p_{F'})(1-f_F)(1-f_{F'})$$
\begin{equation}\label{15b}=\frac{1}{2^3\pi^5}[1+f^0_X(F+F')]\beta(F,F'),
\end{equation}
where incorporated notations \cite{Ignat5}:

\begin{equation}\label{16}\begin{array}{l}
f^0_X(F+F')=\left[\exp\left(-\lambda_F-\lambda_{F'}+E_X/T\right)-1\right]^{-1};\\
\\ \beta(F,F')=\beta(F',F)=\beta(F)+\beta(F');\\
\\
\beta(F)=\displaystyle{\frac{T}{p} \ln
\frac{1+\exp(-\lambda_F+p_+/T)}{1+\exp(-\lambda_F+p_-/T)}-\frac{1}{2}},\\
\end{array}
\end{equation}
 ш
\begin{equation}\label{16a} p_{\pm}=\frac{1}{2}(E\pm p);
\quad E=E_X=\sqrt{m^2_X+p^2}.\end{equation}
Function $\beta(F,F')$ is statistical factor which in Bolzman's
statistics ($\lambda \rightarrow\infty$) is equal to 1. Arguments
$\mathbb{P}$ and $t$ of functions $f^0_X$ ш $\beta$ are dropped
for short. In necessary situations we will write explicitly:
$$f^0_X(x;\mathbb{P},t);\qquad \beta(x,\mathbb{P},t).$$

Subject to (\ref{15a}),(\ref{15b}) equations (\ref{14}) can be
essentially simplified:
\begin{eqnarray}
\frac{d \mathcal{N}_F}{dt}
=a(t){\sum\limits_{X,F}}'\frac{\rho_X}{(2\pi)^3}
|M_{X\rightarrow FF'}|^2 \quad\times\nonumber\\
\label{17}\int\limits_0^{\infty}\frac{p^2[f_X-f^0_X(F+F')]\beta(F,F')
dp}{\sqrt{m_X^2+p^2}}.
\end{eqnarray}
Summation in (\ref{17}) is carrying out by all $X$-bosons,
moreover in quark-quark channels sum of colors of two quarks is
complementary to color of $X$-boson, thus to each type of
$X$-boson are correspond 2 addends in sum.

\subsection{Kinetic equations for X-bosons}
For $X$-bosons optical theorem assumes the form:
\begin{equation}\label{16a}
\sum\limits_{F,F'}d\pi_F d\pi_{F'}
(1-f_F)(1-f_{F'})(W_{x\rightarrow
\bar{F}\bar{F}'}-W_{\bar{X}\rightarrow FF'})=0
\end{equation}
and thus boson equations can be wrote in the form:
\begin{equation}\begin{array}{l}
\sqrt{a^2(t)m^2_X+\mathbb{P}^2}\displaystyle{\frac{\partial f_X}{\partial t}}=\\
\\
\label{18}-\displaystyle{\frac{a(t)}{4\pi}}\sum\limits_{F,F'}|M_{X\rightarrow
\bar{F}\bar{F}'}|^2\beta(F,F')[f_X-f^0_X(F+F')].\\
\end{array}
\end{equation}
Let's draw attention to that important and {\it strict} fact, that
kinematic equations for $X$-bosons become linear differential
equations. Equations for antiparticles receive from (\ref{17})
(\ref{18}) by inversion of overlined indexes of particles -
antiparticles.

If assume that spontaneous violence of $SU(5)$-symmetry happened
at very early stages of cosmological evolution when $X$-bosons
were still ultrarelativistic, then mentioned system of kinetic
equations (\ref{17}), (\ref{18})is necessary to solve with initial
conditions corresponding to initial global thermodynamic
equilibrium and initial baryon and lepton symmetry:
\begin{equation}\label{19}
\left.\lambda_a\right|_{t=0}=0; \qquad f_X(\mathbb{P},0)=f^0_0(0)=
\left[\exp(\mathbb{P}/\mathcal{T}_0)-1\right]^{-1},
\end{equation}
where:
\begin{equation}\label{19a}
\mathcal{T}_0=\mathcal{T}(0)=\displaystyle{\left(\frac{45}{16\pi^3
N}\right)^{1/4}}. \end{equation}

From (\ref{18}) it is obvious that boson functions can be found in
quadratures \cite{Ignat3}, \cite{Ignat4}. These equations in
standard mathematical notation look like:
\begin{equation}\label{19_1}
\dot{f}_i+f_i A_i(t)=Y_i(t),
\end{equation}
point here and further means time derivative, $i,k$ - now indexes
of $X$-bosons, and vectors $A$ ш $Y$ define by correspondences:
\begin{equation}\label{19_2}
A_i(\mathbb{P},t)=\frac{a(t)}{4\pi\sqrt{m^2_ia^2(t)+\mathbb{P}^2_i}}\sum\limits_{k}
|M_{X_i\rightarrow F,F'}|^2\beta(F,F');
\end{equation}
\begin{eqnarray}
Y_i(\mathbb{P},t)=\frac{a(t)}{4\pi\sqrt{m^2_ia^2(t)+\mathbb{P}^2_i}}\times\hskip 2cm\nonumber\\
\label{19_3}\hskip 1cm\sum\limits_{k} |M_{X_i\rightarrow
F,F'}|^2\beta(F,F') f^0_X(F+F'),\label{19_3}
\end{eqnarray}
- summation in (\ref{19_2}),(\ref{19_3}) is carrying out by all
channels of reactions over which given $X$-boson decays.

Solving equations (\ref{19_1}) with initial conditions (\ref{19}),
(\ref{19a}), we will receive solutions for boson functions in
quadratures:
\begin{equation}\label{19_4}
\begin{array}{l}
f_X(\mathbb{P},t)=f_X(\mathbb{P},0) \exp \Bigl(-\int\limits_0^t
A_i(\mathbb{P},t')dt'\Bigr)+\\
\exp \Bigl(-\int\limits_0^t
A_i(\mathbb{P},t')dt'\Bigr)\int\limits_0^t \exp
\Bigl(\int\limits_0^{t'}
A_i(\mathbb{P},t'')dt''\Bigr)Y_i(\mathbb{P},t'')dt''.\\
\end{array}
\end{equation}

Substitution of obtained solution for boson di\-s\-tri\-bu\-ti\-on
functions in kinetic equations for fermions (\ref{17}) leads to
closed system of nonlinear integro-differential equations relative
to chemical potentials.

\subsection{Conservation laws}
In complete kinetic model of Universe it is necessary to add to
equations (\ref{17}), (\ref{18}) another equations which define
evolution of temperature and scale factor. First of these
equations receives by integration of kinetic equations (\ref{9}),
(\ref{10}) with weight $E_i$ and following summation
\cite{Ignat7}:

\begin{equation} \label{20}
\dot{\mathcal{E}}=\dot{a}\sum\limits_X m_X(\mathcal{K}_X +
\mathcal{K}_{\bar{X}}),
\end{equation}
where $\mathcal{E}$ - summary conformal energy density:
\begin{equation} \label{21}
\mathcal{E}=\frac{\pi^2 N' \mathcal{T}^4}{30} +
\\
\sum\limits_X \frac{\rho_X}{2\pi^2}\int\limits_0^{\infty} f_X
\mathcal{P}^2\sqrt{a^2m_X^2+\mathcal{P}^2}d\mathcal{P},
\end{equation}
\begin{equation}\label{22}
\mathcal{K}_X=\frac{a m_X \rho_X}{2\pi^2} \int\limits_0^{\infty}
\frac{f_X\mathcal{P}^2}{\sqrt{a^2m_X^2+\mathbb{P}^2}}d\mathbb{P},
\end{equation}
where $N'$ - number of degrees of freedom of $SU(5)$-model without
accounting of $X$-bosons: $$N'=N-N_X,$$ i.e., in standard
$SU(5)$-model $N'=185-53=132$. And equation that defines evolution
of scale factor is Einstein's equation in which we can put $k=0$,
taking into account that we consider very early Universe $t
\rightarrow 0$:
\begin{equation}\label{23}
\dot{a}^2 = \frac{8\pi}{3}\mathcal{E}.
\end{equation}

Equations (\ref{17}), (\ref{18}), (\ref{20}), (\ref{23}) jointly
with definitions (\ref{21}), (\ref{22}) and initial conditions
(\ref{19}), (\ref{19a}) are the complete system of equations which
describe dynamics of baryogenesis. Let us consider some {\it
strict}  im\-pli\-ca\-ti\-ons of these equations.

\vskip 12pt \noindent {\it 1}.\: Let us suppose that $X$-bosons
lie in thermal equ\-i\-li\-bri\-um, i.e., ignoring kinetic
equations for $X$-bosons (\ref{18}), we will put in (\ref{17})
$$f_X=f^0_X(\lambda_X)=\left[\exp\left(-\lambda_X+\frac{\sqrt{a^2 m^2_X+\mathbb{P}^2}}
{\mathcal{T}}\right)-1\right]^{-1},$$%
where $\lambda_X=-\lambda-\lambda'$. Then exact solution of
equations (\ref{17}), satisfying to initial conditions
(\ref{19}) will be:%
\begin{equation}\label{23a}
\mathcal{N}_a=\const; \qquad \lambda_a=0,
\end{equation}
- i.e., in conditions of statistical equilibrium of $X$-bosons
baryogenesis doesn't execute.

\vskip 12pt\noindent{\it 2}.\: Let us suppose that interactions
$CP$ are invariant, i.e.:
$$W_{X\rightarrow \bar{F},\bar{F}'}=W_{\bar{X}\rightarrow FF'}.$$%
In this situation from (\ref{17}), (\ref{18}) follows again
(\ref{23a}). Thus and in situation of $CP$-invariance of
interactions baryogenesis doesn't execute.

\vskip 12pt\noindent{\it 3}.\:Integrating kinetic equations for
$X$-bosons (\ref{18}) over momentum space and combining this
result with equations (\ref{17}) taking into account initial
conditions (\ref{19}) we will receive two {\it strict}
implications:
\begin{eqnarray}
\sum\limits_{A,C}\left(-\frac{4}{3}\Delta
n_{-4/3}-\frac{1}{3}\Delta
n_{-1/3}\right.+\nonumber\\\label{III.112} +\left.
\frac{2}{3}\Delta n_{2/3}+\frac{2}{3}\Delta n_\alpha
-\frac{1}{3}\Delta n_\kappa -\Delta n_e \right)=0
\end{eqnarray}
- law of conservation of electric charge ($\sum e_a n_a=0$) ш
\begin{eqnarray}
\sum\limits_{A,C}\left[-\frac{2}{3}\left(\Delta n_{-4/3}+\Delta n
_{-1/3}+\Delta n_{2/3}\right)\right. +\nonumber\\\label{III.113}
\left. + \frac{1}{3}\left(\Delta n_\alpha +\Delta n_\kappa\right)
-\Delta n_e -\Delta n_\nu \right]=0
\end{eqnarray}
- law of conservation of difference of baryon and lepton charges
where incorporated notation:
\begin{equation}\label{III.113a}
\Delta n_a = n_a - \bar{n}_a.
\end{equation}
Summation in (\ref{III.112}), (\ref{III.113}) is carrying out by
all charms and colors of particles.%
\vskip 12pt\noindent{\it 4}.\: Supposing now that all $X$-bosons
decayed i.e., proceeding to limit $t\rightarrow \infty$ and
combining co\-r\-re\-s\-pon\-den\-ces (\ref{III.112}),
(\ref{III.113}) with accounting of conditions of symmetry
(\ref{12a}) and notations (\ref{12}), we will receive:

\begin{equation}\label{III.114}
u(\infty)=2d(\infty); \: u(\infty)=2e(\infty)
 \Rightarrow
d(\infty)=e(\infty).
\end{equation}%
From (\ref{III.114}) follows that on completion of decay of
$X$-bosons for each excess charged lepton it will corresponds one
excess katoquark and two excess anoquark, what subject to colors
will give one excess achromic baryon to one lepton. Hence final
baryon charge of Universe will be equal to its lepton charge:
\begin{equation}\label{III.115}
\Delta n_b (\infty)=\Delta n_e(\infty).
\end{equation}
Thus for finding baryon asymmetry of Universe it is enough to
define only one from three values:
$$u(\infty),\quad e(\infty),\quad d(\infty).$$%
\vskip 12pt\noindent{\it 5}.\: In simplest model of baryogenesis
when $CP$-invariance violets only in decay channels of one type of
$X$-bosons, $X_{-4/3}$, from (\ref{III.112}), (\ref{III.113})
strictly follows correspondence in all time of process:
\begin{equation}\label{III.116}
e(t)=d(t),
\end{equation}
and remain only two independent values $u(t)$ ш $d(t)$.

Concrete numerical model of baryogenesis will be published in the
next article.

\subsubsection*{Acknowledgement}
Authors are thankful to D.Y.Ignatyev for translating the paper
into English.

\end{document}